\documentclass[11pt]{article}

 \usepackage{amsmath}
 \usepackage{tabularx}
 \usepackage{array}
 \usepackage{booktabs}
 \usepackage{graphicx}
 \usepackage{longtable}
 \def\CP{{$\cal CP~$}}
 \usepackage{amsfonts}
 \usepackage{float}
\usepackage{hyperref}
\usepackage{bm, geometry, cite, amssymb}
\begin{document}
 
 \title{\LARGE \bf {Bargmann Invariants and Correlated Geometric \CP-Violating Structures in Neutral Meson Systems}}

 \author{
 	Swarup Sangiri \thanks{swarup.phys@gmail.com} \\
   \textit{Physics Department,
	Indian Institute of Technology Kharagpur}\\ \textit{Kharagpur, 721302, India}
}
\date{}
\maketitle

\begin{abstract}
	Bargmann invariants provide a rephasing-invariant description of phase relations among quantum states and offer a geometric perspective on interference phenomena. In this work, we investigate their role in neutral meson systems by constructing cyclic products involving the heavy and light mass eigenstates together with decay-projected states arising from correlated meson decays. Explicit expressions for third-order and fourth-order invariants are obtained in terms of mixing parameters and decay amplitudes. The analysis shows that the associated geometric phases encode \CP-sensitive interference effects between meson-antimeson mixing and decay amplitudes and become trivial in the \CP-conserving limit. Expressing the decay amplitudes in terms of CKM matrix elements reveals quartic combinations with analogous rephasing-invariant weak-phase structure to that of the Jarlskog invariant. 
	
	We further introduce a rephasing-invariant ratio constructed from third- and fourth-order Bargmann invariants, which isolates correlated \CP-violating structures that cannot, in general, be factorized into independent decay-channel contributions and can enhance sensitivity to small deviations from \CP symmetry. The invariants can also be related to parameters governing time-dependent \CP asymmetries in neutral meson decays, thereby providing a geometric interpretation of observable \CP-violating interference effects.
\end{abstract}

\newpage
\section{Introduction}
Understanding phase relations and interference effects is central to the description of quantum systems. In particular, a geometric characterization of such phases provides a complementary perspective to conventional amplitude-based formulations. Bargmann invariants (BIs), introduced as cyclic products of inner products among quantum states \cite{Bargmann1964}, provide a rephasing-invariant framework for describing phase relations and are closely connected to geometric phase concepts such as the Pancharatnam-Berry phase \cite{Pancharatnam1956, Berry1984, AharonovAnandan1987, Rabei1999, Mukunda2003}. Their inherently geometric structure makes them a natural tool for analyzing interference phenomena in systems where phase information plays a crucial role \cite{MukundaSimon1993}.

Neutral meson systems provide a well-established setting in which interference effects arise from the interplay between particle-antiparticle mixing and decay processes \cite{Barr2016}. Systems such as $K^0-\overline{K^0}$, $B_d^0-\overline{B_d^0}$, $B_s^0-\overline{B_s^0}$, and $D^0-\overline{D^0}$ exhibit rich phenomenology associated with mixing and decay dynamics, which are experimentally probed through time-dependent measurements \cite{Branco1999, Buchalla1996}. Entangled quantum states play a fundamental role in describing correlated quantum systems and have been extensively studied in both foundational and experimental contexts \cite{Einstein1935, Schrodinger1935a, Schrodinger1935b, Schrodinger1935c, Bell1964, Horodecki2009}. In particular, neutral meson pairs produced in processes such as $\phi$ and $\Upsilon(4S)$ decays form entangled quantum states, providing a natural setting in which correlations and decay processes can be jointly analyzed \cite{KLOE2006, BABAR2004, Belle2001}. 

A central phenomenon observed in neutral meson systems is \CP violation, which arises from the interplay between mixing and decay processes \cite{Christenson1964, Wolfenstein1964, BigiSanda1981}. \CP violation may appear through different mechanisms, commonly classified as indirect \CP violation associated with the mixing of flavor eigenstates and direct \CP violation originating from complex phases in decay amplitudes \cite{BigiSanda2009, Sarkar2008}. Within the Standard Model, \CP violation originates from complex phases in the Cabibbo-Kobayashi-Maskawa (CKM) quark mixing matrix \cite{Cabibbo1963, KobayashiMaskawa1973}. The strength of \CP violation in the quark sector can be characterized by the rephasing-invariant Jarlskog invariant, which provides a basis-independent measure of \CP violation in weak interactions \cite{Jarlskog1985, Wu1986}. 

While \CP violation in neutral meson systems is conventionally described in terms of decay amplitudes and time-dependent asymmetries, a geometric formulation of the associated phase structure provides a complementary perspective that remains less systematically developed. In this context, BIs provide a natural framework for capturing phase relations associated with interference processes in a basis-independent manner. Geometric phase structures in entangled neutral meson systems have been investigated in recent studies, highlighting their connection to \CP-sensitive phase structures \cite{Sangiri2026}. In contrast, the present approach employs discrete cyclic products of quantum states to characterize interference effects.  

The BI formalism has been previously applied to the neutral kaon system, where a connection between BIs and \CP-violating parameters has been established \cite{SangiriSarkar2023}. In the present work, we extend this approach to a general framework applicable to neutral meson systems by constructing BIs that incorporate both mixing eigenstates and decay-projected states. These decay-projected states arise from entangled meson pairs decaying into specific final states and provide an operational means to encode decay information within the BI framework.

In particular, we analyze third-order and fourth-order BIs constructed from sequences involving heavy and light mass eigenstates together with decay-projected states. These invariants encode interference effects arising from the combined action of meson mixing and decay amplitudes, providing a geometric characterization of \CP-sensitive phase structure. Furthermore, by expressing the decay amplitudes in terms of CKM matrix elements, we obtain quartic combinations that share analogous rephasing-invariant weak-phase structure to that of those appearing in the Jarlskog invariant. This enables a direct comparison between geometric and conventional measures of \CP violation across different neutral meson systems. In addition, this formalism allows the BI structure to be related to experimentally measurable parameters governing time-dependent \CP asymmetries, thereby connecting the geometric foundation with observable quantities.

Building on this structure, we introduce a ratio constructed from third- and fourth-order invariants
\begin{align}
R = \frac{\Delta_4}{\Delta_3(f)\,\Delta_3(g)},\nonumber
\end{align}
where $\Delta_3(f)$ denotes the third-order BI associated with a single decay channel $f$, and $\Delta_4 (\equiv \Delta_4(f,g))$ involves a cyclic sequence incorporating two decay channels $f$ and $g$. This ratio provides additional insight beyond individual invariants, with its structure determined by finite combinations of mixing- and decay-induced interference parameters. While the third-order quantities probe single-channel interference effects, this ratio isolates nontrivial correlations between the \CP-sensitive parameters associated with two distinct decay channels, including phase-phase and mixed interference contributions. As a result, it captures inter-channel geometric phase structures that cannot, in general, be factorized into independent contributions, thereby extending the BI framework to correlated multi-channel interference structures. This construction also facilitates a direct connection with experimentally measurable quantities governing time-dependent \CP asymmetries, strengthening the link between the geometric formulation and observable signatures. These results demonstrate that BIs provide a geometric framework for analyzing both single-channel and correlated \CP-violating interference effects in neutral meson systems. Finally, we briefly comment on the extension of this framework to sequential (cascade) decay processes, where higher-order Bargmann invariants can be constructed to encode correlated interference effects across multiple stages of evolution.

\section{Formalism for Neutral Meson Mixing and Correlated Pairs}
Neutral meson systems such as $K^0-\overline{K^0}$, $B_d^0-\overline{B_d^0}$, $B_s^0-\overline{B_s^0}$, and $D^0-\overline{D^0}$ provide a well-established framework for studying flavor oscillations and \CP violation \cite{PDG2024, Branco1999}. The quantum description of neutral meson mixing is formulated in a two-dimensional Hilbert space $\mathcal{H}$ spanned by the flavor eigenstates $\{|P^0\rangle, |\overline{P^0}\rangle\}$, where $P$ denotes a generic neutral meson. These states $|P^0\rangle$ and $|\overline{P^0}\rangle$ correspond to mesons with definite quark content and are distinguished by flavor quantum numbers such as strangeness, beauty or charm. The internal quantum number remains conserved under strong and QED interactions, while weak interactions can induce particle-antiparticle oscillations, thereby changing the internal quantum number.

Due to the presence of weak interactions, the flavor states are not eigenstates of the effective Hamiltonian governing propagation. The physical propagation eigenstates, commonly referred to as the heavy and light mass eigenstates, can be written as linear combinations of the flavor eigenstates:
\begin{subequations}\label{P_HL}
	\begin{align}
	|P_H\rangle &=p|P^0\rangle +q|\overline{P^0}\rangle, \\
	|P_L\rangle &=p|P^0\rangle -q|\overline{P^0}\rangle,
	\end{align}
\end{subequations}
up to arbitrary phase conventions \cite{BigiSanda2009, Branco1999}.
The parameters $p$ and $q$ encode the mixing between particle and antiparticle states and satisfy the normalization condition
\begin{align}
|p|^2+|q|^2=1.
\end{align}
Indirect \CP violation in the mixing sector manifests when $|q/p|\ne1$, reflecting an asymmetry between the particle and antiparticle components in the propagation level. 

Neutral mesons are often produced in correlated pairs in high-energy experimental environments.  For instance, entangled neutral meson pairs arise in electron-positron collisions at flavor factories through the decays of vector resonances such as $\phi(1020)\rightarrow K^0\overline{K^0}$ and $\Upsilon(4S)\rightarrow B^0\overline{B^0}$ \cite{KLOE2006, BABAR2004}. Because the parent vector resonance has quantum numbers $J^{PC}=1^{--}$, conservation of angular momentum and charge conjugation requires the two-meson system to be produced in an antisymmetric entangled state. The quantum description of such a system requires the tensor-product Hilbert space $\mathcal{H}\otimes\mathcal{H}$, corresponding to the degrees of freedom of the two mesons. The entangled state at production can be expressed as \cite{Lipkin1989, BertlmannGrimus1997, BertlmannHiesmayr2001}
\begin{align}\label{eq:Psi}
|\Psi\rangle=\frac{1}{\sqrt{2}}\left(|P^0\rangle_1|\overline{P^0}\rangle_2-|\overline{P^0}\rangle_1|P^0\rangle_2\right),
\end{align}
where the subscripts label the two mesons. This state configuration implies flavor anticorrelation between the two particles: a measurement projecting one meson onto a definite flavor state immediately determines the flavor of the other. Such correlations provide a direct manifestation of quantum entanglement in the neutral meson system. 

The time evolution of neutral mesons is governed by an effective non-Hermitian Hamiltonian within the Weisskopf-Wigner approximation. In the mass eigenstate basis, the evolution is diagonal, and each state evolves as 
\begin{equation}\label{eq:time_evolution_P_HL}
|P_{H,L}(t)\rangle=e^{-i(m_{H,L}-i{\Gamma_{H,L}\over 2})t}|P_{H,L}(0)\rangle,
\end{equation}
where $m_{H,L}$ and $\Gamma_{H,L}$ denote the masses and decay widths of the heavy and light eigenstates, respectively. This time dependence encodes both oscillatory behavior and exponential decay characteristics of neutral meson systems.  

The decay of neutral mesons into specific final states is described by the transition operator $T$, which represents the weak interaction responsible for the decay process. The decay amplitudes for a final state $f$ are given by \cite{Nir2005} 
\begin{align}\label{eq:decay_amplitude}
A_f=\langle f|T|P^0\rangle, \quad \bar A_f=\langle f|T|\overline{P^0}\rangle.
\end{align}
Here, the transition operator $T$ connects the initial meson state to the multiparticle final state $f$. Differences in magnitude and phase between $A_f$ and $\bar A_f$ signal direct \CP violation in decay amplitudes. Depending on the channel, the final state $f$ may correspond to flavor-specific modes, such as semileptonic decays, or to \CP eigenstates, such as $J/\psi K_S$ \cite{FleischerMannel2001}, which are particularly useful for probing \CP-violating effects. These elements constitute the standard formalism for describing entangled neutral meson systems and provide the foundation for the analysis developed in the following sections. 

\section{Third-Order Bargmann Invariant}
\subsection{Derivation in Terms of  Mixing and Decay Amplitudes}
For three quantum states $\psi_1$, $\psi_2$, $\psi_3$, the third-order Bargmann invariant (BI) is defined by \cite{Bargmann1964, SamuelBhandari1988,  MukundaSimon1993}
\begin{align}
\Delta_3 = (\psi_1,\psi_2) ( \psi_2,\psi_3) (\psi_3,\psi_1),
\end{align}
where $(\psi_i,\psi_j)\equiv\langle \psi_i|\psi_j\rangle$ denotes the inner product of the states $\psi_i$ and $\psi_j$. For normalized pure states, the BI may be equivalently expressed as $\mathrm{Tr}\left(\rho_1\rho_2\rho_3\right)$, where $\rho_i = |\psi_i\rangle\langle\psi_i|$  denotes the density operators corresponding to the participating states, although we will primarily use the overlap representation in this work. This invariant is defined in the projective Hilbert space (ray space) associated with the underlying Hilbert space of the states. To construct the invariant relevant for the correlated neutral meson system, we consider three states belonging to this Hilbert space: $|\psi_1\rangle=|P_H\rangle$, $|\psi_2\rangle=|\psi_f\rangle$, $|\psi_3\rangle=|P_L\rangle$.  The state $|\psi_f\rangle$ arises as the conditional state of the second meson upon decay of the first meson into the final state $f$. The surviving meson is then described by the conditional state
\begin{align}
|\psi_f\rangle_2\propto\langle f|T|\Psi\rangle,
\end{align}
where $T$ denotes the weak transition operator responsible for the decay and acts on the Hilbert space of the decaying (first) meson and $\langle f|$ is defined in that Hilbert space, leaving the second meson in a conditional state. Assuming the initial antisymmetric entangled state of the form in Eq.~(\ref{eq:Psi}) and using the decay amplitudes of Eq.~(\ref{eq:decay_amplitude}), the resulting state of the second meson becomes, up to an overall normalization factor
\begin{align}\label{eq:psif}
|\psi_f\rangle = \frac{1}{\sqrt2} \left( A_f |\overline {P^0}\rangle - \bar A_f |P^0\rangle \right).
\end{align}
Physically, this state represents the flavor superposition of the surviving meson conditioned on the decay of its entangled partner into the channel $f$. The factor $\frac{1}{\sqrt{2}}$ arises directly from the normalization of the initial entangled state in Eq.~(\ref{eq:Psi}); however, no further normalization of $|\psi_f\rangle$ is imposed here, since the BI depends only on the associated rays and is insensitive to overall normalization.

\begin{figure}[H]
	\centering
	\includegraphics[width=0.6\textwidth]{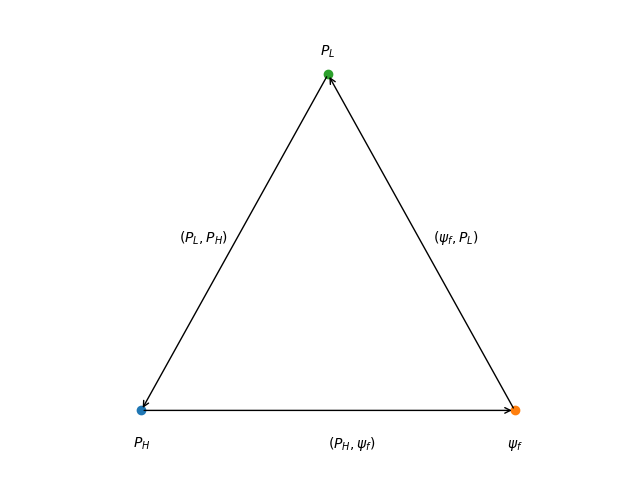}
	\caption{Geometric representation of the third-order Bargmann invariant $\Delta_3$. The closed triangular loop in projective Hilbert space connects the states $P_H \rightarrow \psi_f \rightarrow P_L \rightarrow P_H$. The phase associated with this loop corresponds to the geometric phase $\gamma_{\Delta_3}$.}
	\label{fig:BI3}
\end{figure}

The corresponding BI takes the form
\begin{align}\label{BI3}
\Delta_3=(P_H,\psi_f)(\psi_f,P_L)(P_L,P_H).
\end{align}
As illustrated in Fig.~\ref{fig:BI3}, the third-order BI corresponds to a closed triangular loop in projective Hilbert space.

 This invariant is independent of arbitrary phase choices of the individual states. Under the transformations $|P_H\rangle \to e^{i\chi_1}|P_H\rangle$, 
$|\psi_f\rangle \to e^{i\chi_2}|\psi_f\rangle$, and 
$|P_L\rangle \to e^{i\chi_3}|P_L\rangle$, 
the phases cancel in the cyclic product of overlaps, leaving $\Delta_3$ unchanged. Using the expressions for the mass eigenstates in Eq.~(\ref{P_HL}) and the conditional state in Eq.~(\ref{eq:psif}), the relevant inner products entering the BI can be evaluated directly in the flavor basis. One obtains
\begin{subequations}\nonumber
	\begin{align}
		(P_H,\psi_f) &=\frac{1}{\sqrt{2}}(q^*A_f-p^*\bar A_f),\\
		(\psi_f,P_L) &=\frac{1}{\sqrt{2}}(-qA_f^*-p\bar A_f^*),\\
		(P_L,P_H) &=|p|^2-|q|^2.
	\end{align}
\end{subequations}
Substituting these overlaps in Eq.~(\ref{BI3}) yields
\begin{align}
\Delta_3 = \frac12 (|p|^2-|q|^2) (q^*A_f - p^*\bar A_f) (-qA_f^* - p\bar A_f^*).
\end{align}
Expanding the product yields
\begin{align}\label{eq:BI_3}
\Delta_3 = \frac12(|p|^2-|q|^2) \Big[ |p|^2|\bar A_f|^2 - |q|^2|A_f|^2 + 2i\,\mathrm{Im}(p^*q\,\bar A_f A_f^*) \Big].
\end{align}
This expression separates naturally into real and imaginary parts
\begin{subequations}\label{eq:ReImBIs}
	\begin{align}
	\mathrm{Re}(\Delta_3) &= \frac12(|p|^2-|q|^2) \left( |p|^2|\bar A_f|^2 - |q|^2|A_f|^2 \right),\\
	\mathrm{Im}(\Delta_3) &= (|p|^2-|q|^2)\, \mathrm{Im}(p^*q\,\bar A_f A_f^*).
	\end{align}
\end{subequations}
$\mathrm{Im}(\Delta_3)$ arises from the interference between the mixing parameters and the decay amplitudes and therefore contains the \CP-sensitive phase information of the system.

The  complex phase associated with $\Delta_3$, $\gamma_{\Delta_3}=\arg(\Delta_3)$ has a direct geometric interpretation. In the ray space, it corresponds to the geometric phase accumulated along the closed triangle sequence of rays $P_H\rightarrow\psi_f\rightarrow P_L\rightarrow P_H$. Since BIs depend only on the rays associated with the participating quantum states, the phase $\gamma_{\Delta_3}$ is independent of the normalization of the intermediate states. Using Eq.~(\ref{eq:BI_3}), the geometric phase can be written as 
\begin{align}
\gamma_{\Delta_3} = \arg \Big[ |p|^2|\bar A_f|^2 - |q|^2|A_f|^2 + 2i\,\mathrm{Im}(p^*q\,\bar A_f A_f^*) \Big],
\end{align}
since the overall real prefactor does not affect the phase modulo $\pi$. In regimes corresponding to small \CP violation, $|p|\simeq|q|$, as approximately realized in many neutral meson systems, and the decay amplitudes satisfy  $|A_f|\simeq|\bar A_f|$. In this limit the difference $|p|^2|\bar A_f|^2 - |q|^2|A_f|^2$ is suppressed, while the interference term $\mathrm{Im}(p^*q\,\bar A_f A_f^*)$ is not suppressed at the same parametric order. Consequently, the real part of $\Delta_3$ is parametrically small compared to its imaginary part, and the geometric phase $\gamma_{\Delta_3}$ approaches a maximal value
$\gamma_{\Delta_3}\to\frac{\pi}{2}\mathrm{sgn}\big(\mathrm{Im}(p^*q\,\bar A_f A_f^*)\big)$. Thus the sign of the phase $\gamma_{\Delta_3}$ is directly determined by the \CP-violating interference encoded in $\mathrm{Im}(p^*q\,\bar A_f A_f^*)$.
\subsection{Physical Interpretation and \CP Limit}
The structure of the invariant becomes particularly transparent in several physically relevant limits. In the \CP-conserving case, where the mixing parameters satisfy $|q/p|= 1$, and the decay amplitudes obey $|A_f|=|\bar A_f|$ with aligned phases, the BI vanishes, $\Delta_3=0$. A nonzero imaginary component therefore signals the presence of \CP-sensitive interference effects. 

If indirect \CP violation appears in the mixing sector such that $|p|\ne|q|$, the magnitude of the invariant becomes proportional to the difference $(|p|^2-|q|^2)$. The imaginary component further captures the interference between the mixing phase and the relative decay phase through the quantity $\mathrm{Im}(p^*q\bar A_fA_f^*)$. This term depends on the phase structure $\arg(p^*q)+\arg(\bar A_fA_f^*)$. Thus the BI probes mixing phase, decay phase and their interference.

The behavior also depends on the decay channel considered. For flavor-specific decays, where one of the decay amplitudes vanishes (e.g., $\bar A_f=0$), the BI becomes purely real ($\Delta_3=-{1\over 2}(|p|^2-|q|^2)|q|^2|A_f|^2$) and the geometric phase reduces to a trivial value of $0$ or $\pi$, depending on the sign of $\Delta_3$. In contrast, \CP eigenstate decay channels ($|A_f|\approx|\bar A_f|$) maximize the interference term (${p^*q\bar A_fA_f^*}$) and provide enhanced sensitivity to \CP-dependent geometric effects.

In realistic neutral meson systems the magnitudes of the mixing parameters are typically very close, $|p|\approx|q|$, which suppresses the magnitude of $\Delta_3$.  This observation motivates the consideration of higher-order BIs, which may provide additional \CP-sensitive structures. 
\section{Fourth-Order Bargmann Invariant}
\subsection{Derivation in Terms of  Mixing and Decay Amplitudes}
While the third-order BI represents the simplest nontrivial cyclic phase, higher-order BIs may also be constructed by considering longer cyclic sequences of states. For four states $\psi_1$, $\psi_2$, $\psi_3$, $\psi_4$, the fourth-order BI is defined as
\begin{align}
\Delta_4 &= (\psi_1,\psi_2) (\psi_2,\psi_3) (\psi_3,\psi_4) (\psi_4,\psi_1).
\end{align}
As discussed in the previous section, the invariant depends only on the rays corresponding to the states and is therefore naturally defined in the projective Hilbert space associated with the flavor Hilbert space of the neutral meson system. To construct the fourth-order BI, we consider the cyclic sequence $P_H \rightarrow \psi_f \rightarrow P_L \rightarrow \psi_g \rightarrow P_H$, where $|P_H\rangle$ and $|P_L\rangle$ are the heavy and light propagation eigenstates. The states $|\psi_f\rangle$ and $|\psi_g\rangle$ are given by
\begin{align}
|\psi_f\rangle=\frac{1}{\sqrt2} \left( A_f |\overline{P^0}\rangle - \bar A_f |P^0\rangle \right), \qquad |\psi_g\rangle=\frac{1}{\sqrt2} \left( A_g|\overline{P^0}\rangle - \bar A_g |P^0\rangle \right),
\end{align}
where the factor $\frac{1}{\sqrt{2}}$ is retained from the initial entangled state, and no further normalization is imposed. These states are derived from the antisymmetric entangled state (Eq.~(\ref{eq:Psi})), and denote the conditional states of the surviving meson, following the decay of the entangled partner into the channels $f$ and $g$, respectively. 

The corresponding invariant therefore takes the form
\begin{align}\label{eq:Delta4}
\Delta_4 = (P_H,\psi_f) (\psi_f,P_L) (P_L,\psi_g) (\psi_g,P_H).
\end{align}
\begin{figure}[H]
	\centering
	\includegraphics[width=0.6\textwidth]{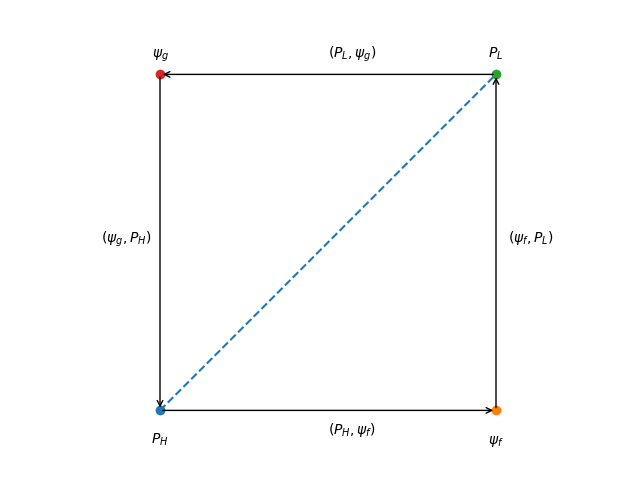}
	\caption{Geometric representation of the fourth-order Bargmann invariant $\Delta_4$. The closed quadrilateral loop involves two decay channels and connects the sequence $P_H \rightarrow \psi_f \rightarrow P_L \rightarrow \psi_g \rightarrow P_H$. This structure encodes interference between mixing and decay processes across multiple channels.}
	\label{fig:BI4}
\end{figure}

The four-state cyclic structure is shown in Fig.~\ref{fig:BI4}, highlighting the role of multiple decay channels. Using the explicit expressions of the eigenstates and the conditional states obtained after the decays into channels $f$ and $g$, one finds
\begin{subequations}\nonumber
	\begin{align}
	(P_H,\psi_f) &=\frac{1}{\sqrt{2}}(-p^*\bar A_f+q^*A_f), \\
	(\psi_f, P_L) &=\frac{1}{\sqrt{2}}(-qA_f^*-p\bar A_f^*), \\
	(P_L,\psi_g) &=\frac{1}{\sqrt{2}}(-p^*\bar A_g-q^*A_g), \\
	(\psi_g, P_H) &=\frac{1}{\sqrt{2}}(qA_g^*-p\bar A_g^*).
	\end{align}
\end{subequations}
Substituting these in Eq.~(\ref{eq:Delta4}) yields
\begin{align}\label{eq:BI_4}
\Delta_4 =\frac{1}{4}\left[ (q^*A_f - p^*\bar A_f) (-qA_f^* - p\bar A_f^*) (-p^*\bar A_g - q^*A_g) (qA_g^* - p\bar A_g^*)\right].
\end{align}
This is expressed entirely in terms of the mixing parameters $p$, $q$ and the decay amplitudes $A_f$, $\bar A_f$, $A_g$, $\bar A_g$. This invariant captures the interference structure associated with a closed sequence of transitions involving heavy-light mixing and two different decay projections. The real and imaginary parts of this invariant can be written as
\begin{subequations}
\begin{align}
	\mathrm{Re}(\Delta_4) &=\frac{1}{4}[(|p|^2|\bar A_f|^2-|q|^2|A_f|^2)(|p|^2|\bar A_g|^2-|q|^2|A_g|^2)]+
	\mathrm{Im}(p^*q\bar A_fA_f^*)\mathrm{Im}(p^*q\bar A_gA_g^*), \\
	\mathrm{Im}(\Delta_4) &=\frac{1}{2}[-(|p|^2|\bar A_f|^2-|q|^2|A_f|^2)\mathrm{Im}(p^*q\bar A_gA_g^*)+
	(|p|^2|\bar A_g|^2-|q|^2|A_g|^2)\mathrm{Im}(p^*q\bar A_fA_f^*)].
 \end{align}
\end{subequations} 
These expressions show that the fourth-order invariant correlates the \CP-sensitive structures associated with two different decay channels. The fourth-order BI therefore provides a rephasing-invariant geometric quantity that links the \CP-sensitive dynamics of two decay channels within the entangled neutral meson system.

The geometric phase associated with the fourth-order BI is given by
\begin{align}
\gamma_{\Delta_4}=\arg(\Delta_4).
\end{align}
Using Eq.~(\ref{eq:BI_4}) we obtain
\begin{align}
 \gamma_{\Delta_4}= \arg \left[ (q^*A_f - p^*\bar A_f) (-qA_f^* - p\bar A_f^*) (-p^*\bar A_g - q^*A_g) (qA_g^* - p\bar A_g^*) \right].
\end{align}
This phase represents the geometric phase accumulated when the system undergoes the cyclic sequence of state overlaps
$P_H \rightarrow \psi_f \rightarrow P_L \rightarrow \psi_g \rightarrow P_H$. The geometric phase therefore extends this dependence to two decay channels, thereby encoding inter-channel interference. 
\subsection{Physical Interpretation and \CP Limit}
The physical content of the BI becomes transparent when considering several limiting cases. First, consider the situation in which the two decay channels are identical, $f=g$. In this case one obtains
$A_g=A_f$, $\bar A_g=\bar A_f$,
and the invariant reduces to
\begin{align}
\Delta_4 = \frac{1}{4} |qA_f^*-p\bar A_f^*|^2 |qA_f^*+p\bar A_f^*|^2 .
\end{align}
Since this quantity is real and positive,
$\arg(\Delta_4)=0$.
Thus the geometric phase becomes trivial when both decay channels coincide, indicating that two distinct decay projections are required in order to generate a nontrivial four-state interference loop.

Next consider the \CP-conserving limit. \CP symmetry implies
$|q/p|=1$, and for \CP eigenstate decay channels $\bar A_f = \eta_f A_f$, $\bar A_g = \eta_g A_g $,
where $\eta_f $ and $\eta_g $ are the \CP eigenvalues of the final states. Substituting these relations into the invariant yields a purely real $\Delta_4$. Therefore a nonvanishing imaginary part of the invariant provides a rephasing-invariant probe of \CP-violating interference between mixing and decay processes.

Another instructive limit arises in the absence of mixing, where the mass eigenstates coincide with the flavor eigenstates,
$|P_H\rangle=|P^0\rangle$, $|P_L\rangle=|\overline{P^0}\rangle$ .
The invariant then reduces to
\begin{align}
\Delta_4 = \frac{1}{4}\bar A_fA_f^*A_g\bar A_g^*.
\end{align}
The phase of this depends only on the relative phases of the decay amplitudes. Thus the invariant no longer probes mixing-decay interference, demonstrating that heavy-light mixing is essential for generating the geometric phase associated with the cyclic sequence considered here.

In principle, BIs of arbitrary order may be constructed from longer cyclic sequences of states. The $n^{\mathrm{th}}$-order BI is given by
\begin{align}
\Delta_n=(\psi_1,\psi_2)(\psi_2,\psi_3)...(\psi_{n-1},\psi_n)(\psi_n,\psi_1).
\end{align}
However a fundamental property of geometric phases in projective Hilbert space is that any $n$-point BI for $n\ge 4$ can be reduced to $\Delta_3$'s and $\Delta_2$'s \cite{Mallesh2012}.

Thus, while fourth- or higher-order BIs can be reduced to lower order BIs at the geometric level, the corresponding invariants in physical systems can encode additional correlations. In particular, the four-point invariant constructed above captures the interference loop involving two distinct decay channels. This fourth-order BI therefore provides a complementary geometric probe of the interplay between heavy-light mixing and decay processes in the neutral meson system.

\section{Connection to Quark-Level \CP Invariant}
The BIs derived in the previous sections contain \CP-sensitive contributions arising from the interference between decay amplitudes and meson mixing. In this section, we express the decay amplitudes in terms of CKM matrix elements and examine the resulting quartic combination, thereby making the connection with the fundamental source of \CP violation in the Standard Model explicit.

From Eq.~(\ref{eq:ReImBIs}), the \CP-sensitive contribution arises from the imaginary component, governed by the combination $\mathrm{Im}(p^*q\bar A_fA_f^*)$.
To make the connection with the underlying quark-level origin of \CP violation explicit, the decay amplitudes, within the weak effective Hamiltonian formalism, can be expressed in terms of CKM matrix elements \cite{Buras1998, Beneke2000}. We write 
\begin{subequations}
	\begin{align}
	A_f  &= \sum_{\alpha} V_{\alpha i}V_{\alpha j}^*\,T_{\alpha}^{(f)}, \\
	\bar A_f  &= \sum_{\alpha} V_{\alpha i}^*V_{\alpha j}\,T_{\alpha}^{(f)},
	\end{align}
\end{subequations}
where $T_{\alpha}^{(f)}$ denote the corresponding hadronic matrix elements which contain the strong-interaction dynamics, $\alpha=u,c,t$ and $i,j=d,s,b$. Substituting these expressions into the \CP-sensitive term gives
\begin{align}
\bar A_f A_f^*=\sum_{\alpha,\beta}\left( V_{\alpha i}^*V_{\alpha j}V_{\beta i}^*V_{\beta j} \right)T_{\alpha}^{(f)}{T_{\beta}^{(f)}}^*.
\end{align}
 Consequently, the \CP-sensitive contribution of BI becomes
\begin{align}
\mathrm{Im}(p^*q\bar A_f A_f^*) = \sum_{\alpha,\beta}  \mathrm{Im}\left( p^*q V_{\alpha i}^*V_{\alpha j}V_{\beta i}^*V_{\beta j} T_{\alpha}^{(f)}{T_{\beta}^{(f)}}^*\right).
\end{align}
The weak interaction phases therefore appear through quartic products of CKM matrix elements of the form 
\begin{align}\label{weak_phase}
V_{\alpha i}^*V_{\alpha j}V_{\beta i}^*V_{\beta j}. 
\end{align}
Such quartic CKM products are structurally similar to the rephasing-invariant combinations that define the Jarlskog invariant $J=\mathrm{Im}(V_{\alpha i}V_{\beta j}V_{\alpha j}^*V_{\beta i}^*) $ responsible for \CP violation in the Standard Model \cite{Jarlskog1985}. Although the combination appearing here is not identical to the standard Jarlskog form, it contains analogous weak phase structure associated with \CP-violation through products of four CKM matrix elements. This establishes a bridge between geometric phase structures defined at the meson level and \CP violation at the quark level.

An important feature of these quartic combinations is that they are invariant under quark field rephasing transformations
\begin{align}
V_{\alpha i}\rightarrow e^{i\phi_\alpha}V_{\alpha i}e^{-i\chi_i}\nonumber,
\end{align} 
under independent rephasing of up-type and down-type quark fields. Under this transformation the product $V_{\alpha i}^*V_{\alpha j} V_{\beta i}^*V_{\beta j}$ remains unchanged. As a result, the imaginary part of the expression in Eq.~(\ref{weak_phase}) represents a physically meaningful quantity that is independent of arbitrary quark field phase conventions. This demonstrates that the \CP-sensitive contribution encoded in the BI probes analogous rephasing-invariant weak phases that govern \CP violation at the quark level. 

A similar analysis may be performed for the fourth-order BI. In this case, the cyclic product involves two decay channels, $f$ and $g$, and the resulting expression contains two independent \CP-sensitive combinations of mixing and decay amplitudes. When expressed in terms of the underlying weak transition amplitudes, each of these contributions leads to quartic products of CKM matrix elements, analogous to the structure obtained for the third-order invariant. Consequently, the third- and fourth-order BIs encode \CP-sensitive interference through Jarlskog-type combinations involving four CKM elements.

The magnitude and phase of these quartic CKM combinations depend on the quark transitions governing the decay amplitudes of the neutral meson under consideration \cite{PDG2024}. In kaon decays involving $s\rightarrow d$ transitions the relevant CKM matrix elements are primarily $V_{us}$, $V_{ud}$, $V_{cs}$ and $V_{cd}$. In the standard phase convention these elements are predominantly real, and therefore the imaginary part of the corresponding quartic combinations is strongly suppressed, which is consistent with the observation that \CP violation in the neutral kaon system arises predominantly through mixing effects \cite{Buras1999, Littenberg1993}. 

In contrast, neutral $B_d$ meson decays involve transitions such as $b\rightarrow c\bar{c}s$ or $b\rightarrow u\bar{u}d$, where CKM matrix elements from different generations contribute simultaneously. The resulting quartic products contain sizable complex phases originating from the CKM matrix. Consequently, the imaginary parts of these combinations can be comparatively significant, giving rise to \CP-violating interference effects observed in time-dependent $B$-meson decay asymmetries \cite{CarterSanda1981, Sanda1985}. 

For the $B_s$ system, the relevant CKM factors lead to a weak phase that is small within the Standard Model, implying that the imaginary part of the corresponding quartic combinations is suppressed. This is reflected in the small value of the mixing-induced \CP-violating phase measured in $B_s$ decays \cite{LHCb_2013, LHCb_2019}. Similarly, in the $D^0$ system, the dominant transitions involve first and second generation quarks whose CKM matrix elements carry negligible complex phases, leading to highly suppressed \CP-violating contributions \cite{Grossman2007}. As a result, the system dependence of the quartic CKM structure reflects the hierarchical pattern of weak phases present in the CKM matrix.   

 \section{Geometric \CP Observables and Correlated Interference Structures in Neutral Mesons}
 In this section, we relate the BIs to experimentally accessible quantities and introduce a derived ratio that captures correlated \CP-violating structures across multiple decay channels. The BIs derived in Eq.~(\ref{eq:BI_3}) depend on the neutral-meson mixing parameters $p$ and $q$ together with the decay amplitudes $A_f$ and $\bar A_f$. These amplitudes correspond to experimentally measurable transition matrix elements, $A_f=\langle f|T|P^0\rangle$, $\bar A_f=\langle f|T|\overline{P^0}\rangle$, which enter the standard description of neutral-meson mixing and decay \cite{PDG2024, BigiSanda2009}. In experimental analysis, the interaction between meson-antimeson mixing and decay into a common final state $f$ is commonly characterized by the parameter \cite{Nir2015}
\begin{align}
\lambda_f = \frac{q}{p}\frac{\bar A_f}{A_f}.
\end{align}
The \CP-sensitive component of $\Delta_3$ involves the structure $p^*q\bar A_f A_f^*$. This combination can be written directly in terms of experimentally determined quantities. Writing
\begin{align}
p^*q\bar A_f A_f^* = |A_f|^2\,p^*q\frac{\bar A_f}{A_f},
\end{align}
and using the definition of $\lambda_f$, 
$\frac{\bar A_f}{A_f}=\frac{p}{q}\lambda_f $,
one obtains, up to an overall phase convention for the mixing parameters
\begin{align}\label{eq:Im_part}
p^*q\bar A_f A_f^* = |A_f|^2 |p|^2\lambda_f .
\end{align}
This relation expresses the BI contribution in terms of the rephasing invariant quantity $\lambda_f$. Consequently, the third-order BI derived in Eq.~(\ref{eq:BI_3}) can be rewritten as
\begin{align}
\Delta_3=\frac{1}{2}(|p|^2-|q|^2)|A_f|^2\left[{|p|^4\over |q|^2}|\lambda_f|^2-|q|^2+2i|p|^2\mathrm{Im}(\lambda_f)\right],
\end{align}
and the \CP-odd component entering the BI becomes
\begin{align}
\mathrm{Im}(p^*q\bar A_f A_f^*) = |p|^2 |A_f|^2\mathrm{Im}(\lambda_f).
\end{align}
This suggests that the \CP-sensitive part of the BI is governed by the same parameter $\mathrm{Im}(\lambda_f)$ that controls the experimentally measured time-dependent \CP asymmetry \cite{CarterSanda1981, BigiSanda2009}
\begin{align}
A_{CP}(t) = \frac{\Gamma(\overline{P^0}(t)\to f)-\Gamma(P^0(t)\to f)} {\Gamma(\overline{P^0}(t)\to f)+\Gamma(P^0(t)\to f)} .
\end{align}
The coefficient of the oscillatory term in this asymmetry is proportional to
\begin{align}
S_f=\frac{2\,\mathrm{Im}(\lambda_f)}{1+|\lambda_f|^2},
\end{align}
indicating that the geometric phase associated with BI depends on the same underlying interference parameter measured in time-dependent \CP violation studies of neutral mesons. The invariant therefore gives a geometric representation of the \CP-violating interference between meson mixing and decay amplitudes. 

Extending this connection to multiple decay channels, we introduce the following ratio constructed from third- and fourth-order invariants
\begin{align}
R = \frac{\Delta_4}{\Delta_3(f)\Delta_3(g)}.
\end{align}
For notational simplicity, the dependence of $\Delta_4$ on the pair of decay channels $(f,g)$ is left implicit. This ratio isolates genuine inter-channel \CP-sensitive correlations that are not accessible through individual third-order invariants. 
To evaluate this explicitly, we use the expressions obtained for the third-order invariants in Eq.~(\ref{eq:BI_3}),
\begin{subequations}
	\begin{align}
	\Delta_3(f) = \frac{1}{2} (|p|^2 - |q|^2)(a_f + i b_f),\\
	 \Delta_3(g) = \frac{1}{2} (|p|^2 - |q|^2)(a_g + i b_g),
	\end{align}
\end{subequations}
where
\begin{align}
a_i = |p|^2|\bar A_i|^2 - |q|^2|A_i|^2, \qquad b_i = 2\mathrm{Im}(p^*q\bar A_i A_i^*), \quad (i=f,g).
\end{align}
The product of the two third-order invariants then becomes
\begin{align}
\Delta_3(f)\Delta_3(g) 
&= \frac{1}{4} (|p|^2 - |q|^2)^2 \left[(a_f + i b_f)(a_g + i b_g)\right]\nonumber\\
&=\frac{1}{4} (|p|^2 - |q|^2)^2 \left[(a_f a_g - b_f b_g) + i(a_f b_g + a_g b_f)\right].
\end{align}
On the other hand, the fourth-order invariant derived in Eq.~(\ref{eq:BI_4}), can be written in the form
\begin{align}
\Delta_4 = \frac{1}{4} \left[a_f a_g + b_f b_g + i(a_g b_f - a_f b_g)\right].
\end{align}
Substituting these expressions into the definition of $R$, we obtain
\begin{align}
R =\frac{a_f a_g + b_f b_g + i(a_g b_f - a_f b_g)}{(|p|^2 - |q|^2)^2 \left[a_f a_g - b_f b_g + i(a_f b_g + a_g b_f)\right]}.
\end{align}
This expression depends only on combinations of the quantities $a_i$ and $b_i$, which are themselves determined by the mixing parameters and decay amplitudes. Expressing these in terms of the experimentally accessible parameters $\lambda_f$ and $\lambda_g$, all overall amplitude normalization factors cancel provided a consistent normalization of decay amplitudes is used across channels, so that the ratio $R$ depends only on rephasing-invariant combinations of $\lambda_f$ and $\lambda_g$, as shown in the following explicit expression
\begin{equation}
R=
\frac{
	\begin{aligned}
	& \left(\frac{|p|^4}{|q|^2}|\lambda_f|^2-|q|^2\right)
	\left(\frac{|p|^4}{|q|^2}|\lambda_g|^2-|q|^2\right)
	+4|p|^4\mathrm{Im}(\lambda_f)\mathrm{Im}(\lambda_g) \\
	& + 2i|p|^2\left[
	\mathrm{Im}(\lambda_f)\left(\frac{|p|^4}{|q|^2}|\lambda_g|^2-|q|^2\right)
	-\mathrm{Im}(\lambda_g)\left(\frac{|p|^4}{|q|^2}|\lambda_f|^2-|q|^2\right)
	\right]
	\end{aligned}
}{
	\begin{aligned}
	& (|p|^2-|q|^2)^2 \Bigg[
	\left(\frac{|p|^4}{|q|^2}|\lambda_f|^2-|q|^2\right)
	\left(\frac{|p|^4}{|q|^2}|\lambda_g|^2-|q|^2\right)
	-4|p|^4\mathrm{Im}(\lambda_f)\mathrm{Im}(\lambda_g) \\
	& \quad +2i|p|^2\left[
	\mathrm{Im}(\lambda_g)\left(\frac{|p|^4}{|q|^2}|\lambda_f|^2-|q|^2\right)
	+\mathrm{Im}(\lambda_f)\left(\frac{|p|^4}{|q|^2}|\lambda_g|^2-|q|^2\right)
	\right]
	\Bigg]
	\end{aligned}
}\,.
\end{equation}
It is important to note that the ratio $R$ becomes ill-defined in the exact \CP-conserving limit, where $|p|^2 = |q|^2 $ and consequently $\Delta_3(f) = \Delta_3(g) = 0$. This behavior is not pathological but rather reflects the fact that the third-order invariants vanish when \CP symmetry is exact, eliminating the underlying interference structure required for defining the ratio. The quantity $R$ is therefore meaningful only in the presence of \CP violation, where it serves as a sensitive probe of deviations from the \CP-conserving limit.

In the regime of small \CP violation, the quantities $a_f$, $a_g$ remain nonvanishing in general, while the factor $(|p|^2-|q|^2)^2$ in the denominator becomes parametrically small. Consequently, the ratio $R$ exhibits an enhancement proportional to
\begin{align}
\frac{1}{(|p|^2-|q|^2)^2}\nonumber,
\end{align}
making it particularly sensitive to small deviations from \CP symmetry in the mixing sector. The divergence of $R$ reflects the vanishing of the denominator in the \CP-conserving limit and indicates that the ratio becomes highly sensitive to small deviations from \CP symmetry. In this sense, $R$ provides a sensitive diagnostic of \CP violation, enhancing otherwise suppressed interference contributions arising from neutral-meson mixing.

In order to make the parametric dependence of the ratio $R$ more transparent in the regime of small \CP violation, it is useful to perform a systematic expansion in terms of physically small quantities. This allows one to disentangle the contributions arising from mixing-induced, direct, and interference \CP violation, and to clarify the origin of the enhancement observed in $R$. We parametrize \CP violation in mixing as
\begin{align}
|p|^2 =\frac{1}{2}(1+\epsilon_m), \qquad |q|^2 = \frac{1}{2}(1-\epsilon_m), \qquad |\epsilon_m|\ll 1,
\end{align}
so that $\epsilon_m = |p|^2 - |q|^2$ provides a convenient measure of the deviation from \CP symmetry. This parameter is directly related to the standard observable $|q/p|$ via
\begin{align}
\left|\frac{q}{p}\right|^2=\frac{1-\epsilon_m}{1+\epsilon_m}\approx1-2\epsilon_m.
\end{align}
The direct \CP violation in decay can be expressed as
\begin{align}
|\bar A_i|^2 = |A_i|^2 (1+\delta_i), \qquad |\delta_i|\ll 1,
\end{align}
which follows from defining $\delta_i$ as the fractional difference between particle and antiparticle decay rates. The interference parameter is expressed as
\begin{align}
\lambda_i=|\lambda_i|e^{i\phi_i}, \qquad \mathrm{Im}(\lambda_i)\equiv\beta_i,
\end{align}
which is directly related to experimentally measured time-dependent asymmetries. It is important to note that $\beta_i = \mathrm{Im}(\lambda_i)$, with $\lambda_i = \frac{q}{p}\frac{\bar A_i}{A_i}$, is not strictly independent of the small parameters $\epsilon_m$ and $\delta_i$, since both the magnitude and phase of $\lambda_i$ receive contributions from mixing and decay. However, $\beta_i$ is primarily controlled by the phase of $\lambda_i$, and therefore need not be small even when $\epsilon_m \ll 1$ and $\delta_i \ll 1$. Accordingly, in the expansion we treat $\beta_i$ as an independent parameter and retain all terms involving $\beta_i$, without assigning it a definite scaling relative to the other small quantities. The implications of different parametric regimes for $\beta_i$ are analyzed separately below.

 With these definitions, the quantities entering the invariants take the form
\begin{subequations}
	\begin{align}
	a_i = |A_i|^2 \left( \epsilon_m + \frac{\delta_i}{2} + \frac{\epsilon_m \delta_i}{2} \right),\\
	b_i = |A_i|^2 \left( \beta_i + \epsilon_m \beta_i \right),
	\end{align}
\end{subequations}
where the expression for $b_i$ follows from Eq.~(\ref {eq:Im_part}).

Substituting these expressions into the definition of $R$, and retaining terms up to quadratic order in the small parameters $\epsilon_m$ and $\delta_i$, one obtains
\begin{equation}\label{eq:R_NLO}
R= \frac{1}{\epsilon_m^2}\cdot\frac{
	\begin{aligned}
	&
	\beta_f\beta_g(1+2\epsilon_m+\epsilon_m^2)+\left(\epsilon_m+\frac{\delta_f}{2}\right)\left(\epsilon_m+\frac{\delta_g}{2}\right)\\
	&+i\left[(\beta_f-\beta_g)(\epsilon_m+\epsilon_m^2)+(\beta_f\delta_g-\beta_g\delta_f)\left(\frac{1}{2}+\epsilon_m\right)\right]
	\end{aligned}
}{
    \begin{aligned}
    &
    -\beta_f\beta_g(1+2\epsilon_m+\epsilon_m^2)+\left(\epsilon_m+\frac{\delta_f}{2}\right)\left(\epsilon_m+\frac{\delta_g}{2}\right)\\
    &+i\left[(\beta_f+\beta_g)(\epsilon_m+\epsilon_m^2)+(\beta_f\delta_g+\beta_g\delta_f)\left(\frac{1}{2}+\epsilon_m\right)\right]
    \end{aligned}
}.
\end{equation}
In performing the expansion of the ratio $R$, we treat $\epsilon_m$ and $\delta_i$ as parametrically small quantities and retain terms up to quadratic order, including $\mathcal{O}(\epsilon_m^2)$, $\mathcal{O}(\delta_i^2)$, and mixed terms of the form $\mathcal{O}(\epsilon_m \delta_i)$. No assumption is made regarding the magnitude of the interference parameters $\beta_i = \mathrm{Im}(\lambda_i)$, which are therefore kept to all orders. This corresponds to an expansion in $\epsilon_m$ and $\delta_i$ only, while treating the phase-sensitive parameters $\beta_i$ exactly.
The structure of $R$ contains several qualitatively distinct types of correlations, together with higher-order correlations built from the same underlying combinations. These include phase–phase correlations $\beta_f \beta_g$, mixing-interference correlations $\epsilon_m\beta_i$,
mixing–direct correlations $\epsilon_m \delta_i$, and direct–direct correlations $\delta_f \delta_g$,
together with antisymmetric combinations such as
$\beta_f - \beta_g$, $\delta_g \beta_f - \delta_f \beta_g$,
which encode genuine inter-channel differences in \CP-violating phases.

Keeping terms up to leading nontrivial order in $\epsilon_m$ and $\delta_i$, while allowing arbitrary dependence on $\beta_i$, the ratio in Eq.~(\ref{eq:R_NLO}) takes the form
\begin{align}\label{R_LO}
R=\frac{1}{\epsilon_m^2}\cdot\frac{\beta_f\beta_g(1+2\epsilon_m)+i\left[\epsilon_m(\beta_f-\beta_g)+\frac{1}{2}(\delta_g\beta_f-\delta_f\beta_g)\right]}{-\beta_f\beta_g(1+2\epsilon_m)+i\left[\epsilon_m(\beta_f+\beta_g)+\frac{1}{2}(\delta_g\beta_f+\delta_f\beta_g)\right]}.
\end{align}
This expression makes explicit that, at leading nontrivial order, the observable is governed by an interplay between phase correlations $(\beta_f\beta_g)$ and interference terms involving mixing and direct \CP violation. To further clarify its behavior, it is useful to analyze limiting regimes depending on the relative size of the interference parameters $\beta_i = \mathrm{Im}(\lambda_i)$.

If the interference parameters dominate over the small expansion parameters, i.e. $\beta_i \gg \epsilon_m, \delta_i$, the dominant contribution arises from the phase–phase correlation term $\beta_f \beta_g$. Dividing the numerator and denominator of Eq.~(\ref{R_LO}) by $\beta_f \beta_g$, one obtains
\begin{align}
R = \frac{1}{\epsilon_m^2} \cdot \frac{
	(1+2\epsilon_m)
	+ i\left[
	\epsilon_m\!\left(\frac{1}{\beta_g} - \frac{1}{\beta_f}\right)
	+ \tfrac{1}{2}\!\left(\frac{\delta_g}{\beta_g} - \frac{\delta_f}{\beta_f}\right)
	\right]
}{
	-(1+2\epsilon_m)
	+ i\left[
	\epsilon_m\!\left(\frac{1}{\beta_g} + \frac{1}{\beta_f}\right)
	+ \tfrac{1}{2}\!\left(\frac{\delta_f}{\beta_f} + \frac{\delta_g}{\beta_g}\right)
	\right]
}.
\end{align}

In this form, the dominant structure is manifestly controlled by the constant term $(1+2\epsilon_m)$, while all subleading contributions enter through suppressed ratios such as $\epsilon_m/\beta_i$ and $\delta_i/\beta_i$.
This demonstrates that, in this regime, the observable $R$ is predominantly sensitive to correlated \CP-violating phases across different decay channels. The expansion is stable, and the antisymmetric combinations appearing in the imaginary part provide a direct measure of inter-channel differences in the interference parameters. In this sense, $R$ acts as a clean probe of phase-driven \CP violation.

When the interference parameters are of the same order as the small expansion parameters, i.e. $\beta_i \sim \epsilon_m, \delta_i
$, no single contribution dominates. The phase–phase term $\beta_f \beta_g$, the mixing–interference terms $\epsilon_m \beta_i$ and the direct–interference terms $\delta_i \beta_j$ all contribute at comparable order.
In this regime, the structure of $R$ reflects a genuine interplay between different sources of \CP violation. The observable becomes sensitive to correlated combinations such as
$\beta_f \beta_g$, $ \delta_f \beta_g$, $ \delta_g \beta_f$,
and can exhibit nontrivial cancellations depending on the relative signs and magnitudes of these terms. As a result, the interpretation of $R$ is less straightforward, but it encodes richer information about the joint structure of mixing and decay dynamics.

In the limit where $\beta_i$ becomes parametrically small, the interference contributions are suppressed. Retaining the leading contributions after suppressing the quadratic phase-phase term $\beta_f\beta_g$, one obtains
\begin{align}
R = \frac{1}{\epsilon_m^2} \cdot \frac{
	i\left[\epsilon_m(\beta_f - \beta_g) + \tfrac{1}{2}(\delta_g \beta_f - \delta_f \beta_g)\right]
}{
	i\left[\epsilon_m(\beta_f + \beta_g) + \tfrac{1}{2}(\delta_f \beta_g + \delta_g \beta_f)\right]
},
\end{align}

showing that the observable becomes dominated by subleading interference effects.
In the strict limit $\beta_f = \beta_g = 0$, all terms involving the interference parameters vanish. In this case, one must return to the quadratic contributions retained in the full expansion, yielding
$R = \frac{1}{\epsilon_m^2}$.
This limiting behavior is physically significant: it corresponds to the absence of \CP-violating phases in the interference parameter $\lambda_i$, so that the ratio $R$ loses all sensitivity to inter-channel correlations and reduces to a purely mixing-driven enhancement. The nontrivial structure of $R$ therefore originates entirely from the presence of nonzero $\mathrm{Im}(\lambda_i)$.

Overall, this systematic expansion and regime analysis demonstrate that the ratio $R$ provides a unified framework for probing correlated \CP-violating effects. Its behavior interpolates between a phase-dominated regime, where it cleanly measures interference between decay channels, and a magnitude-dominated regime, where it becomes sensitive to the interplay between different small \CP-violating parameters.

Thus, while the third-order invariants $\Delta_3(f)$ and $\Delta_3(g)$ probe single-channel interference effects, the ratio $R$ isolates genuine inter-channel correlations between the \CP-sensitive parameters $\lambda_f$ and $\lambda_g$. This correlation structure cannot, in general, be factorized into  independent contributions from individual decay modes and therefore represents a qualitatively new feature of the geometric formulation. 

Since the quantities $p$, $q$, $A_f$, and $\bar A_f$ are determined from time-dependent decay measurements and global fits, such as those performed by the CKMfitter Group and the UTfit Collaboration \cite{CKMfitter2023, UTfit2004}, the invariants $\Delta_3$, $\Delta_4$, and the ratio $R$ can, in principle, be constructed from experimentally extracted parameters. The ratio $R$, in particular, provides a geometrically motivated observable that captures correlated \CP-violating structures across multiple decay channels in a manifestly rephasing-invariant manner.

The geometric framework developed here admits a natural extension to more complex processes involving sequential neutral meson mixing, such as cascade decays of the form $B \to D \to f$, where interference arises from the combined effects of two distinct mixing sectors \cite{Shen2024}. In such systems, the total amplitude is built from successive time evolutions and decay projections, and can therefore be interpreted in terms of extended sequences of state overlaps in projective Hilbert space.
Within the present framework, this extension can be incorporated by iterating the construction of conditional states. In particular, one may consider the situation where the conditional state $|\psi_f\rangle$, defined in Eq.~(\ref{eq:psif}), subsequently evolves and contributes to a second decay channel $g$. This effectively leads to a cascade decay process, in which the surviving meson undergoes sequential mixing and decay transitions.

Without introducing additional notation, such a scenario corresponds to a modified conditional state that encodes contributions from both decay stages. A BI can then be constructed in direct analogy with Eq.~(\ref{BI3}) by replacing $|\psi_f\rangle$ with this cascade-modified state $|\psi_f^{(g)}\rangle$, yielding a closed overlap structure in projective Hilbert space analogous to the third-order construction
\begin{align}
\Delta_3=(P_H, \psi_f^{(g)})(\psi_f^{(g)},P_L)(P_L,P_H).
\end{align} 
While the effective state $|\psi_f^{(g)}\rangle$ encodes the combined evolution through the intermediate decay, more generally one may construct higher-order Bargmann invariants for cascade processes, allowing the exploration of correlated interference effects across sequential transitions. The associated geometric phase would therefore incorporate the combined effects of the two mixing sectors within a unified overlap structure.

In contrast to the single-decay case, one expects that the resulting invariant contains interference terms involving contributions from both decay amplitudes, reflecting the correlated nature of the sequential process. In particular, the \CP-sensitive phase encoded in the imaginary part of the invariant is expected to receive contributions from both stages of the decay chain, thereby providing a geometric realization of correlated \CP-violating effects arising from sequential mixing sectors. A detailed analysis of these structures, including their explicit form and phenomenological implications, is left for future work.

\section{Summary}
In this work, we investigated the role of Bargmann invariants in characterizing \CP-sensitive phase structures in neutral meson systems. The analysis focuses on cyclic products of states involving the heavy and light mass eigenstates together with decay-projected states. Although the BIs themselves are constructed from single-meson states, the entangled two-meson state enters through the decay-projected states $\psi_f$ and $\psi_g$, which encode correlations between mixing and decay processes. 

For the third-order BI constructed from the sequence $|P_H\rangle$, $|\psi_f\rangle$, and $|P_L\rangle$, the explicit calculation demonstrates that the \CP-sensitive structure arises from two key contributions: the difference $(|p|^2-|q|^2)$, which reflects \CP violation in neutral-meson mixing, and the interference combination $p^*q\bar A_f A_f^*$, which encodes the phase relation between mixing parameters and decay amplitudes. The resulting geometric phase is therefore directly tied to the interplay of these two effects and becomes trivial in the \CP-conserving limit.

The fourth-order BI extends this construction to two decay channels, leading to analogous \CP-sensitive structures involving $p^*q\bar A_f A_f^*$ and $p^*q\bar A_g A_g^*$. This higher-order structure naturally incorporates multi-channel interference and provides access to correlations that are not directly visible at the level of individual decay modes.

Expressing the decay amplitudes in terms of CKM matrix elements shows that these contributions involve quartic CKM products containing analogous rephasing-invariant weak-phase structure to that of the Jarlskog invariant. The system dependence of the quartic CKM structure reflects the hierarchical structure of weak phases across different neutral meson systems.

To further isolate these correlated effects, we introduced the ratio $R = \frac{\Delta_4}{\Delta_3(f)\Delta_3(g)}$,
constructed from third- and fourth-order invariants. While the third-order quantities probe single-channel interference, this ratio captures nontrivial inter-channel correlations between the \CP-sensitive parameters associated with different decay modes. In particular, it separates correlated contributions from factorized single-channel effects and provides a rephasing-invariant characterization of multi-channel geometric phase structure. The behavior of $R$ also reflects the interplay between direct and indirect \CP violation, exhibiting increased sensitivity in regimes where mixing-induced \CP violation is small, while becoming ill-defined in the exact \CP-conserving limit where the underlying geometric phase structure becomes trivial. In addition, a systematic expansion of $R$ in terms of small mixing- and decay-induced \CP-violating parameters allows one to disentangle phase-phase, mixing-interference, mixing-direct, and direct--direct correlation structures, while treating the interference parameters $\mathrm{Im}(\lambda_i)$ without imposing a fixed parametric hierarchy.

Finally, the interference structure $p^*q\bar A_f A_f^*$ can be expressed in terms of the experimentally accessible parameter $\lambda_f=(q/p)(\bar A_f/A_f)$, which governs the time-dependent \CP asymmetry in neutral meson decays. This connection extends naturally to the multi-channel case through the ratio $R$, thereby providing a geometric framework in which both single-channel and correlated \CP-violating observables can be interpreted in terms of Bargmann invariants. The framework also admits a natural extension to sequential (cascade) decay processes, where higher-order invariants can encode interference effects across multiple stages of evolution.

{\it\bf Acknowledgement :} I wish to thank Prof. Utpal Sarkar and Prof. Arghya Taraphder for support and encouragement. I would like to thank MoE, Government of India for research fellowship.

\end{document}